# A study of thermally-induced optical bistability and the role of surface treatments in Si-based mid-infrared photonic crystal cavities

**Raji Shankar,[1,*] Irfan Bulu,[1] Rick Leijssen,[2] Marko Lončar[1]**

[1]*School of Engineering And Applied Sciences, Harvard University, Cambridge, MA 02138, USA*
[2]*Huygens Laboratory, Leiden University, P. O. Box 9504, 2300 RA Leiden, Netherlands*
[*]*shankar2@fas.harvard.edu*

**Abstract:** We report the observation of optical bistability in Si-based photonic crystal cavities operating around 4.5 µm. Time domain measurements indicate that the source of this optical bistability is thermal, with a time constant on the order of 5 µs. Quality (Q) factor improvement is shown by the use of surface treatments (wet processes and annealing), resulting in an increase of Q-factor from 11,500 to 29,300 at 4.48 µm. After annealing in a $N_2$ environment, optical bistability is no longer seen in our cavities.

---

## References and links

---

## 1. Introduction

Optical bistability in Si photonic crystal microcavities is a well-known phenomenon at telecommunications wavelengths[1-5]. Since the refractive index of Si can change either directly or indirectly due to incident light intensity, the resonance wavelength of a microcavity also changes with a buildup of optical power in the cavity. This results in a positive feedback process which allows the cavity to act as a bistable switch, with an off-resonance, or “empty” state and an on-resonance, or “loaded” state. The power dependence of the refractive index can be attributed to various effects, including the $\chi3$ of Si, free-carrier dispersion, and the thermo-optic effect. Strong light confinement in nano-photonic devices at telecom wavelengths (e.g. 1.55 µm), and in high quality (Q) factor photonic crystal cavities in particular, results in two-photon absorption processes which lead to a pronounced thermo-optic effect due to free-carrier absorption. The high-Q factors and low mode volumes of photonic crystal microcavities lead to a low bistability threshold, with a switching energy that scales roughly with $V/Q^2$. This limits the operation of Si-based photonic devices to low powers at the telecom wavelength range. However, it has recently been proposed [6-7] and demonstrated [8-9] that Si devices operating at longer, mid-infrared (IR) wavelengths would not suffer from this problem due to the lack of two-photon absorption effects. Therefore, Si devices operating in the mid-IR could be of interest for the realization of high-power optical interconnects, as well as enable nonlinear wavelength conversion and amplification of optical signals directly in Si. This, along with standard mid-IR applications such as trace-gas sensing and optical wireless, is the main reason for the recent interest in Si mid-IR photonics [6-14].

We recently demonstrated the operation of photonic crystal cavities in an air-bridged Si membrane platform at 4.4 µm[13]. In [13], we noted the likely presence of optical bistability in our cavities at high input powers. Here, we present an in-depth study of this effect and investigate the origin of the observed nonlinearity using time domain measurements (Section 2). Our results indicate that the observed bistability is thermal (thermo-optic effect) in nature (Section 3). We also explore the effects of standard microelectronic treatments and annealing on the bistability and Q-factors of our cavities (Section 4). These processes allow us to mitigate the bistability in our cavities and achieve a Q-factor of 29,300, which to our knowledge is the highest Q-factor measured in silicon mid-infrared optical cavities to date.

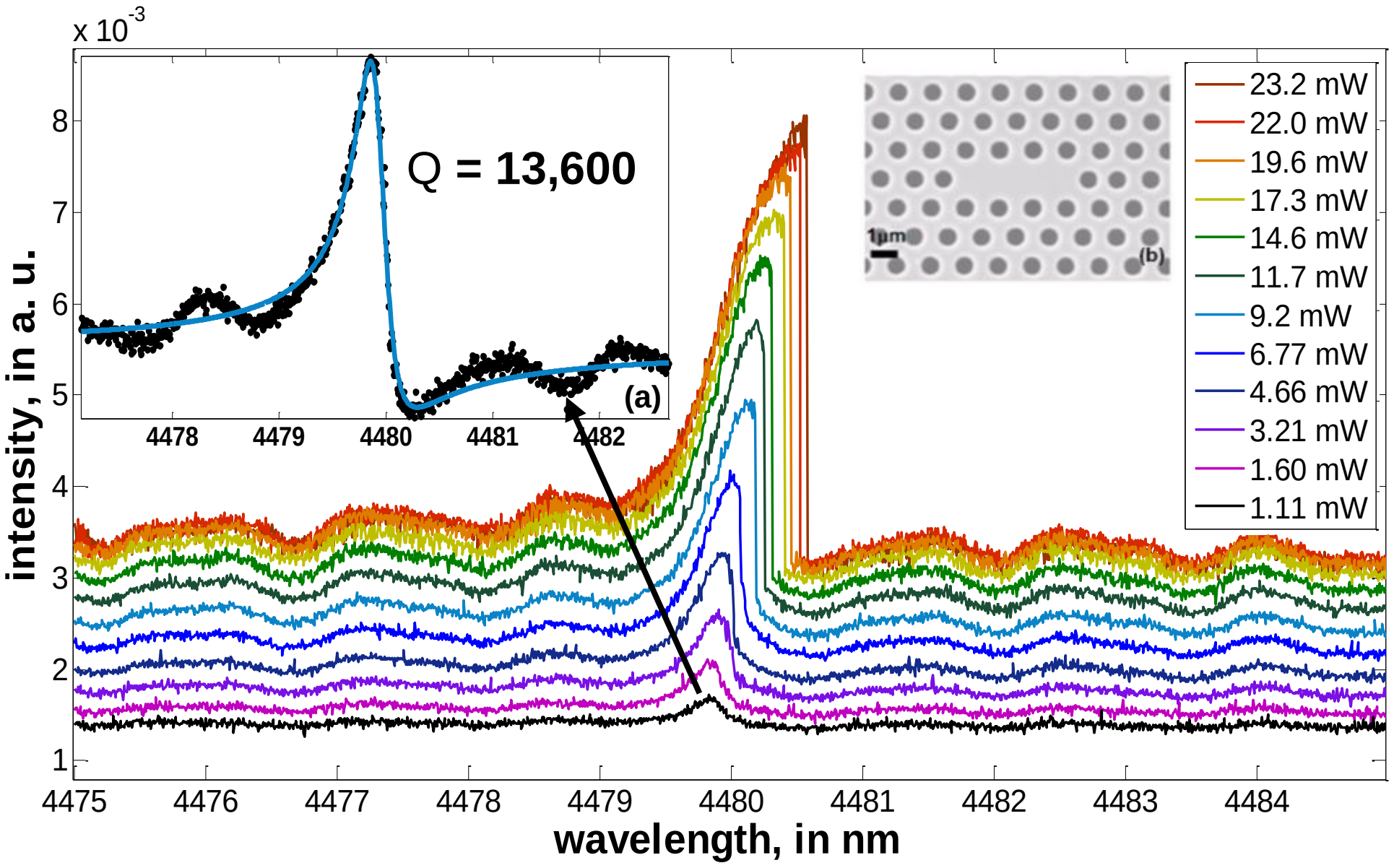

Figure 1. Cavity spectrum taken at various input powers showing characteristic bistable lineshape. Power levels given represent the power after the objective. Inset (a) shows cold cavity resonance and Fano fit to lineshape; Q = 13,600. Inset (b) shows scanning electron micrograph of one of our L3 cavities.

## 2. Fabrication and Characterization

Our devices are L3 photonic crystal cavities made in an air-bridged silicon membrane platform, with device thickness $t$ = 500 nm (Figure 1). The basic fabrication process is detailed in [13]. The hole periodicity is $a$=1.31 µm, and the radius is 0.259$a$ = 339 nm. Cavities with varying hole shifts $s$ were fabricated, with the highest Q of 13,600 being measured in an as-processed cavity with $s$ = 0.15$a$ at 4479.80 nm.

We used the well-known resonant scattering method to couple light into our cavities via free-space[13, 15]. Light from a tunable quantum cascade laser (QCL) with emission from 4.315 to 4.615 µm (Daylight Solutions, Inc.) is sent into a ZnSe objective lens (N.A. = 0.22) and focused onto the sample, which is placed so that the cavity mode polarization is oriented at 45° with respect to the E-field of the laser spot.

Figure 1 shows the spectrum of one of our mid-IR cavities (with $s$ = 0.15$a$), at various input powers, with the wavelength swept from shorter to longer wavelengths. A variable attenuator, placed between the laser output and the input polarizer, was used to vary the power incident on the cavity. The incident power incident on the cavity was calibrated using a thermal power meter placed after the ZnSe objective lens. The evidence of bistability can be clearly seen at input powers over 3 mW, with the characteristic asymmetric bistable lineshape. We note, however, that the coupling efficiency of our setup is less than 20% and therefore we estimate that less than 0.6 mW of incident optical power is actually coupled into the cavity. It is also clear that in our mid-IR cavities, the bistability is due to a redshift of cavity resonance with increasing power, which as we will discuss later, narrows the possible origins of the optical bistability. The characteristic bistable lineshape is due to the fact that as the cavity is swept through red detunings from the cold cavity resonance (4479.80 nm): the cavity remains in a loaded state, because of positive feedback from the power-dependent refractive index, until the detuning becomes so large that the cavity transitions sharply into the empty state. As the input power increases, this effect is more dramatic, and the wavelength at which the transition occurs red-shifts. At powers below the bistability threshold of 3.2 mW, the lineshape becomes more Fano-like[16]. Using a Fano fit, we measured the Q-factor of this cavity using the spectrum taken at the lowest power, obtaining a value of 13,600.

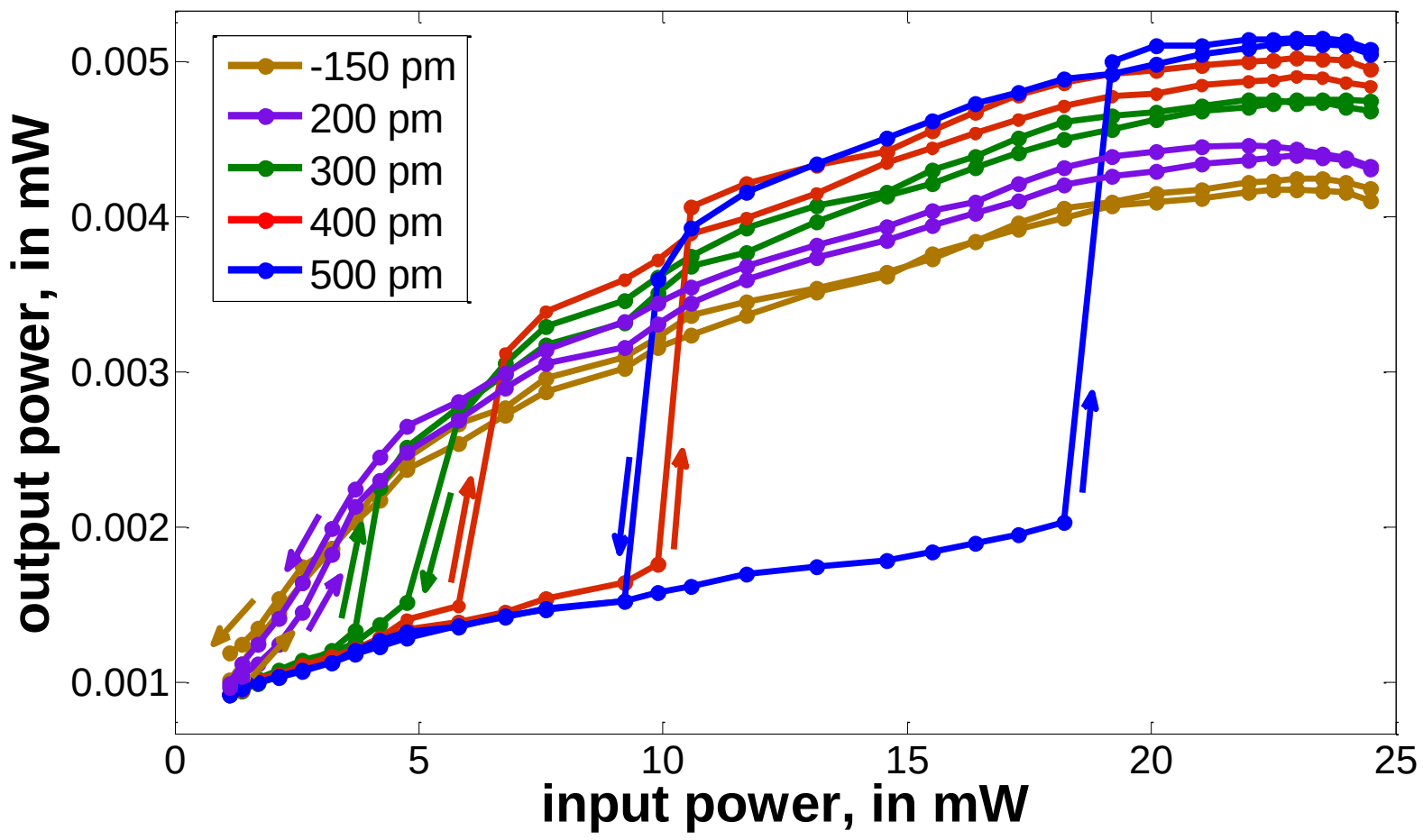

Figure 2. Power input-power output hysteresis curves. Onset of bistability is seen at a detuning of 300 pm.

To further study the observed bistability, we generated hysteresis curves by setting the laser at a fixed red detuning from the cold cavity resonance and measuring the input power and output power for upward and downward sweeps of input lower power level (Figure 2)[2, 17-18]. At detunings $\delta \geq 300$ pm, clear bistability is seen, with the hysteresis growing with increasing $\delta$. Again, we can see from the figure that the bistability threshold is a little over 3 mW, which is consistent with Figure 1.

## 3. Time Domain Measurements

A variety of mechanisms can alter the effective refractive index as a function of the intensity stored inside a Si cavity. They can generally be classified into three categories: thermal effects, free carrier dispersion, and $\chi 3$ effects[18-19]. Thermal effects occur when light is absorbed into the cavity, changing the temperature of the cavity and hence the refractive index through the thermo-optic effect, resulting in an increased refractive index and redshift of the cavity resonance. Different absorption mechanisms that can lead to the thermo-optic effect include the intrinsic absorption of silicon in mid-IR (phonon-assisted absorption), various surface absorption effects, free-carrier absorption, absorption from native oxide formed on Si surfaces, and multiphoton absorption. A comparison of the intrinsic absorption of Si, free-carrier absorption and native oxide absorption is given in Table 1.

Table 1. Comparison of potential sources of absorption in our Si photonic crystal cavities at 4.5 µm.

| **Source of Absorption** | **α (cm$^{-1}$)** | **$Q_{absorption}$** |
|---|---|---|
| Linear absorption of Si [6] | 0.004 | $1 \times 10^7$ |
| Free carrier absorption (n-type doping, ρ=50 Ω·cm) [20] | 0.002 | $2.4 \times 10^7$ |
| Native oxide growth (4 nm) [21] | 9.8 | $1.8 \times 10^6$ |

Multiphoton absorption is unlikely at this wavelength, as at least 4 photons are needed to overcome the Si bandgap energy. Free carrier absorption, either from carriers generated via multiphoton absorption or from carriers introduced by doping of the wafer is also unlikely: (i) our highly resistive Si wafer ($\rho$=50 $\Omega\cdot$cm) has a loss constant of $\alpha < 0.002$ cm$^{-1}$ at 4.5 µm [20], which cannot explain the amount of heating that we observe (see below); (ii) free-carrier generation by multi-photon (four-photon) effects is unlikely. Free carrier dispersion results in a blue-shift of cavity resonance and thus cannot be the cause of our bistability. Therefore, we conclude that among the different possible absorption mechanisms, intrinsic Si absorption, surface effects, and absorption due to the thin native oxide layer could explain our results. In addition, direct nonlinear processes due to the $\chi$3 of Si (the Kerr effect in particular) could explain our results. However, the Kerr effect results in an instantaneous change of refractive index, whereas thermally-induced refractive index changes occur on a much slower time scale (on the order of µs). Therefore, time domain analysis can help us separate these effects and establish whether or not our nonlinearity is primarily thermal [18-19, 22].

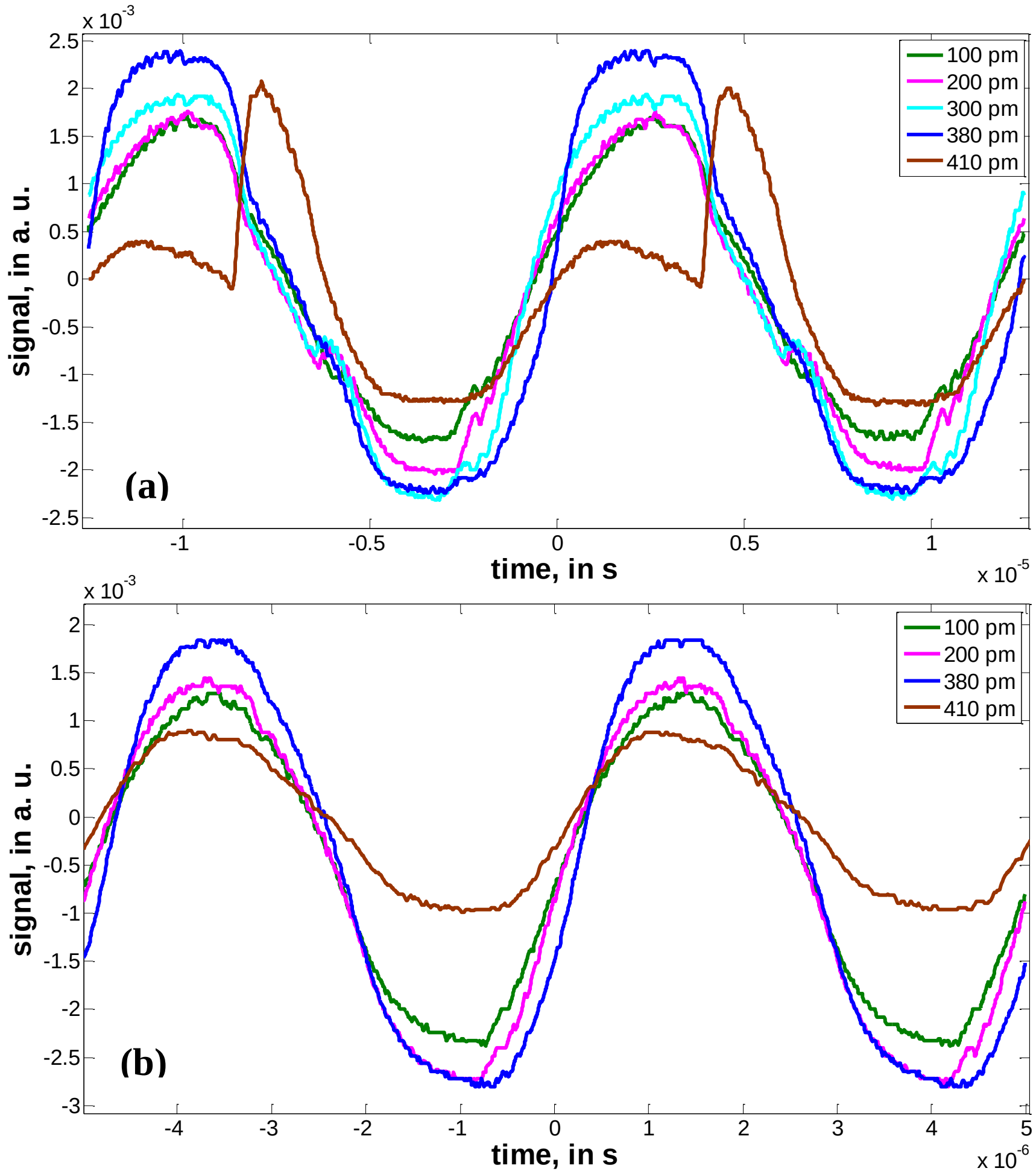

Figure 3. Temporal response of bistable cavities excited with light modulated at 80kHz and 200kHz. (a) For 80 kHz modulation, clear bistability can be observed for detunings higher than 300 pm: the waveform deviates from the sinusoid, taking on more of a square waveform at detunings of 300 and 380 pm. At $\delta$ = 410 pm, we see a sharp discontinuity in the waveform because we are at the drop-off wavelength at this particular power level. (b) For 200 kHz modulation, distortion is reduced and detected waveform tends to sinusoidal.

To do this, we followed the method of [18] and [22], in which the input laser signal is modulated with a sine wave before being coupled into our cavity. In the bi-stable regime the output signal should be distorted, and tend towards a square wave [18], due to the hysteresis loop (Figure 2). The minimum detuning at which bistability can be observed can be estimated as $\delta_{\min} = \frac{\sqrt{3}}{2} FWHM$, where FWHM (full-width half maximum) is the linewidth of the cavity resonance. In our experiments, we explored a range of detunings δ, both above and below the theoretical minimum $\delta_{min}$.=285 pm. In Figure 3, we show the results of this experiment at 80 kHz (Figure 3a) and 200 kHz (Figure 3b) modulation frequency. For δ<285pm at 80 kHz, no effect of bistability is seen in Figure 3(a), and a sinusoidal waveform is recovered at the output. However, at higher detunings, we can see a clear deviation from the sinusoid, with δ = 300 pm and δ = 380 pm resulting in a square-looking waveform. At δ = 410 pm, we are at the drop-off wavelength for this particular input power-level and hence we recover a sharp discontinuity in the waveform. In contrast, when the input signal is modulated at 200 kHz (Figure 3(b)), a much less distorted, sinusoidal-like, waveform is recovered for both δ = 380 pm and δ = 410 pm. This indicates that our bistability is slow in nature, and therefore is due to thermal effects and not instantaneous χ3 effects. We note that 200 kHz was the modulation frequency at which all bistability effects disappeared, and therefore we estimate the thermal time constant to be about 5 µs. This is also consistent with our finite-element modeling of thermal effects in our cavities.

## 4. Effects of Micro-electronic Treatments on Device Performance and Bistability

In order to identify the impact of different surface effects on absorption, we performed various microelectronic treatments which alter the surface properties of our cavities. First, an additional HF dip was performed to remove the native oxide formed on our Si cavities due to prolonged exposure to air (several days). We estimated that 3-4 nm of native oxide could be formed on our cavities [23]. However, since $SiO_2$ has a high material absorption in the mid-IR (α = 9.8 $cm^{-1}$ at 4.5 um[21]), even 4 nm can have a detectable absorptive effect, with a calculated $Q_{absorption}$ of $1.8x10^6$. We carried out a brief HF dip (10 seconds) and then transferred the sample to a $N_2$ purged environment to minimize oxidation and the effects of environmental moisture. The results at low excitation power are shown in Figure 4(b), with the pre-HF dip cavity spectrum shown in Figure 4(a). As a result of the HF dip, the Q increased from from 11,500 to 21,000 and blueshifted about 500 pm. This is a much larger increase in Q than we expected by only taking into account native-oxide removal, and indicates that the $Q_{absorption}$ of the removed material ≈ 25,000 (nearly two orders of magnitude smaller than estimated). We speculate that this discrepancy may be due to differences in absorbance of native oxide (with a rough surface) and bulk measurements taken in silica glass [21]. In addition, various surface-states formed on native oxide may explain such unexpected increase in Q. However, at higher powers, bistability was still clearly present in the cavity (data not shown). Any potential change in bistability threshold was difficult to quantify, since different amounts of power couple into the cavity during different experiments.

In addition to the effects of native oxide, we were concerned about the effects of surface absorption states in the oxide-silicon interface[24-25] and roughness[24]. To address this, we performed a repeated piranha clean (3:1 $H_2SO_4$:$H_2O_2$)/HF dip cycle as proposed by Borselli *et al.* in [24] to decrease surface roughness and other surface absorption effects. Each cycle consisted of 10 minutes of piranha clean, followed by a 3X de-ionized (DI) water rinse (30 seconds each), then 1 minute of 10:1 HF acid dip, followed by a 2X DI water rinse. This cycle was performed three times in total. The purpose of this cyclic process is to oxidize the surface and sidewalls of the cavities through the piranha clean and then remove the oxide using HF, resulting in less surface roughness and fewer surface absorption states. After this process, the cavities were placed in a $N_2$ purged environment and characterized. As shown in Figure 4(c), the Q-factor of the cavity went up to 22,700 and the cavity resonance blueshifted by about 18 nm. We estimate $Q_{absorption}$ of the removed material to be $2.8x10^5$. Using finite-difference time-domain (FDTD) methods, we calculated that an 18 nm blueshift meant that the Si device layer had decreased in thickness by about 4 nm. Again, even after this processing step, the bistability was still present at high powers.

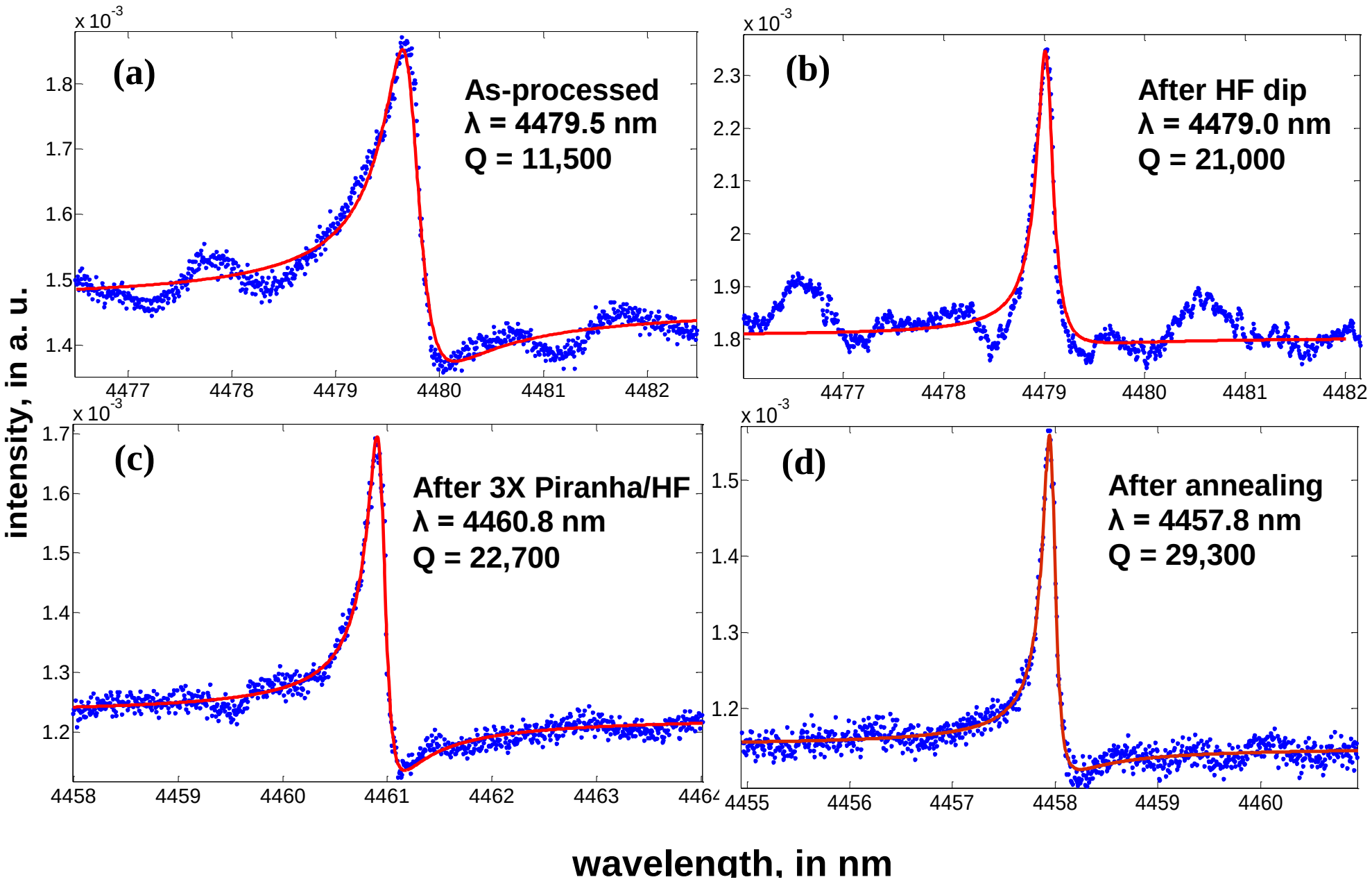

Figure 4. Effect of different post-fabrication treatments on cavity Q. All spectra are taken at low pump powers, well below bistability threshold. (a) Cavity as-processed. (b) After HF dip. (c) After 3X Piranha/HF cycle. Note large blue-shift. (d) After annealing.

Finally, we decided to investigate the effects of annealing in order to remove water moisture from our devices. We annealed our samples by ramping up from room temperature to 500°C over a period of 5 hours, holding the temperature at 500°C for 2 hours, and then ramping down to room temperature for one hour. The Q-factor of the cavity went up even further after this treatment, to 29,300, and the resonance wavelength blue-shifted by about 3 nm, as shown in Figure 4(d). $Q_{absorption}$ of the removed material corresponded to about $1x10^5$. After this first annealing step, the bistability did not disappear at high input powers. However, the increase in Q-factor was significant, indicating to us that water moisture likely has a considerable absorptive effect at 4-5 μm.

A further literature review indicated annealing in a $N_2$ environment can be used to desorb hydrogen from the surface of Si [26]. Si-H bonds can absorb energy in the 250-300 meV range [26], which corresponds roughly to the resonance wavelength of our cavities. We returned to our cavities after a month's time for another cycle of surface treatments (Figure 5) in order to address this potential source of absorption. The Q-factor of the cavity decreased to about 22,000 in this period. As before, wavelength spectra were taken before and after each step (not shown). Re-doing the HF dip and Piranha/HF cycling resulted in an increase of Q-factor (to 25,000) and a shift in resonance wavelength from 4458 nm to 4431 nm due to the material removed through this process. Bistability was still present in the cavity after these two steps were completed. However, after repeating the anneal, but this time flowing $N_2$ through the chamber, we noticed that the cavity spectrum no longer had the characteristic bistable lineshape (blue curve, taken before annealing) and instead was more Fano-like (red curve) at similar input powers. To confirm that the cavity was no longer bistable, we compared input power vs. output power hysteresis loops taken before and after the anneal (insets). The detuning for both hysteresis loops was kept constant, at a value of $\delta$ = 330 pm, which is approximately $2\delta_{min}$ for a cavity with Q ≈ 25,000. The pre-anneal hysteresis loop (blue) clearly shows the bistable turn-on and turn-off, whereas the post-anneal hysteresis loop (red) lacks these sharp bistable transitions. Some hysteresis is present in the post-anneal, indicating that some small sources of absorption still remain in the cavity [2], but bistability is no longer seen after annealing treatment. The Q-factor did not change measurably after annealing, reaching a value of 25,600. After the sample was placed in air for a few days, the cavity spectrum returned to its pre-annealed state, with bistability appearing at the same input powers as before. However, additional annealing in a $N_2$ environment caused the bistability to disappear again. Therefore, based on our results we concluded that water moisture and/or Si-H bonds on the surfaces of the silicon device layer is the primary source of absorption that led to bistability.

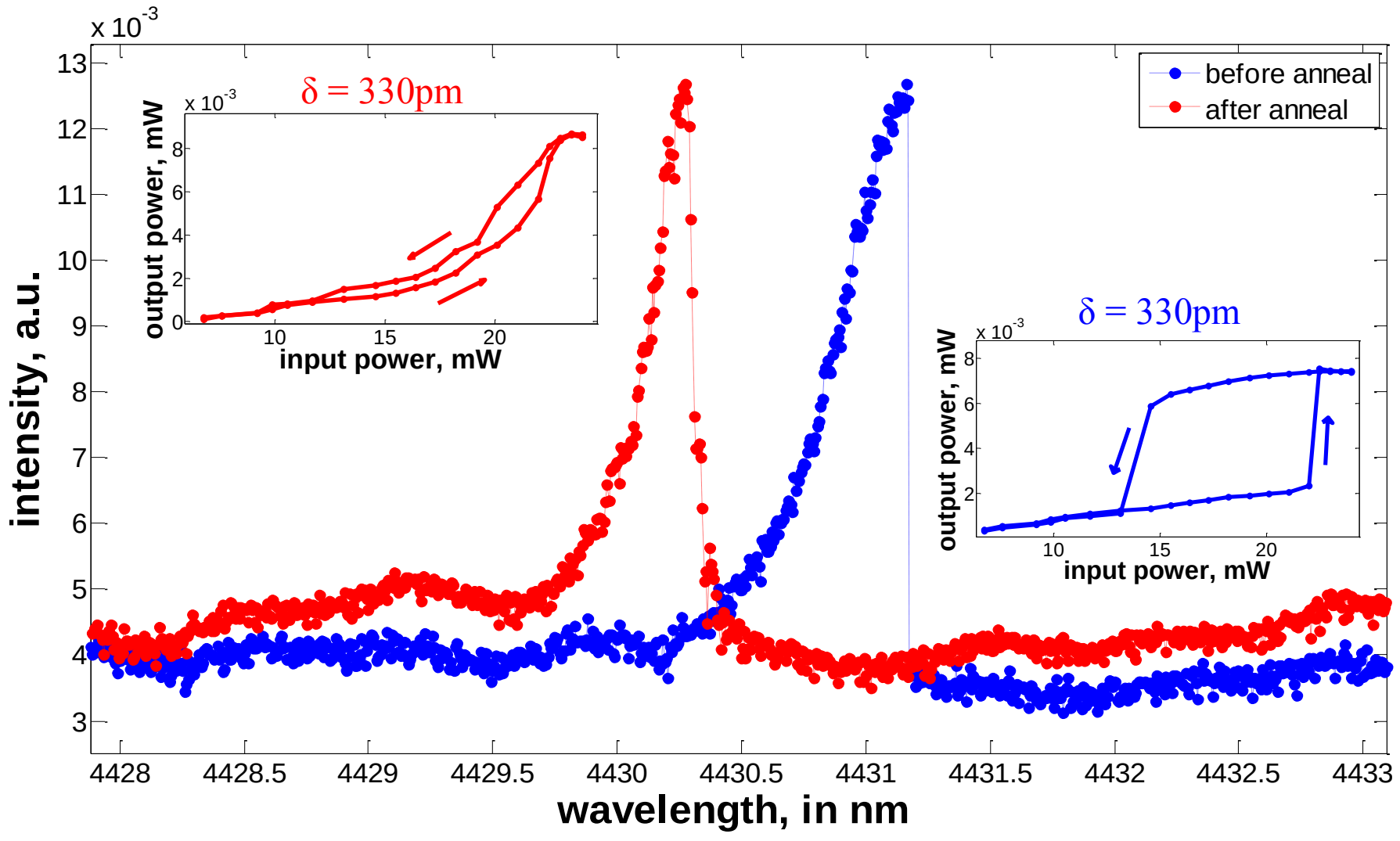

Fig 5. Cavity spectrum taken before (blue) and after (red) annealing in a $N_2$ environment. The blue spectrum and corresponding hysteresis loop (taken at $\delta$ = 330 pm $\approx 2\delta_{min}$) show clear evidence of bistability, while the red spectrum and corresponding hysteresis loop (taken at the same detuning of $\delta$ = 330 pm) indicate that bistability is no longer present.

Though our main objective in performing these microelectronic treatments on our cavities was to elucidate the source of our bistability, we also discovered a method by which to substantially improve the quality factor of our cavities, from an initial Q of 11,500 to 29,300. We believe that at this point the Q of our cavities is limited by the design; the Q-factor obtained from FDTD calculations was 34,000. These micro-electronic treatments and annealing processes can be applied to future mid-infrared Si devices to help improve performance and make the goal of achieving high-Q (Q > 100,000) resonators in the mid-IR more achievable.

## 5. Conclusions

In conclusion, we have observed the presence of bistability in Si mid-IR photonic crystal cavities at 4.5 µm. Through the use of time-domain measurements, we have established that the bistability we see in our mid-infrared photonic crystal cavities is thermal in nature. Finally, we explored the effects of various post-processing methods (HF dips, piranha/HF cycles, and annealing) on the bistability and Q-factors of our cavities. These methods resulted in an increase of Q from 11,500 to 29,300 after processing. Further investigation led to the discovery that annealing in a $N_2$ environment removed bistability from our cavities even at the highest pump powers possible in our setup, pointing to water moisture or Si-H bonds at the surface as the likely cause of our bistability. Surface effects play a large role in microstructured devices in thin Si device layers, since they have a much higher surface-to-volume ratio as compared to bulk crystalline Si. Because of this, the surface treatments we investigated will likely be necessary to achieve nonlinear optics at the 4-5 µm range [7-8], allowing us to pump high powers (necessary in order to see nonlinear effects) into Si cavities without deleterious effects. Decreasing the thermal resistance of the structure will also help mitigate the effects of bistability through the use of different cavity geometries or material platforms, such as silicon-on-sapphire[12]. Alternatively, optical bistability has been shown to have a number of applications in the telecom [2-3] so the possibility of transferring these applications to the mid-IR exists.

**Acknowledgements**

The authors would like to thank Parag Deotare and Qimin Quan for many helpful discussions, and Ian Burgess for his help with annealing. This work is supported in part by the NSF CAREER grant (ECCS-0846684) and generous support from Schlumberger-Doll Research Center (Cambridge, MA). Device fabrication was performed at the Center for Nanoscale Systems at Harvard University. R. S. would like to thank the NSF GRFP.